\documentclass[8.5pt,twoside,twocolumn]{article}
\usepackage[super,sort&compress,comma]{natbib} 
\usepackage{mhchem}
\usepackage{times,mathptmx}
\usepackage{sectsty}
\usepackage{balance} 
\usepackage{amssymb}
\usepackage{enumerate}
\usepackage{graphicx} 
\usepackage{lastpage}
\usepackage[format=plain,justification=raggedright,singlelinecheck=false,font=small,labelfont=bf,labelsep=space]{caption} 
\usepackage{fancyhdr}
\def\lsim{\mathrel{\rlap{\lower4pt\hbox{\hskip1pt$\sim$}}
    \raise1pt\hbox{$<$}}}         
\def\gsim{\mathrel{\rlap{\lower4pt\hbox{\hskip1pt$\sim$}}
    \raise1pt\hbox{$>$}}}         

\begin{document}

\thispagestyle{plain}
\fancypagestyle{plain}{
\renewcommand{\headrulewidth}{1pt}}
\renewcommand{\thefootnote}{\fnsymbol{footnote}}
\renewcommand\footnoterule{\vspace*{1pt}%
\hrule width 3.4in height 0.4pt \vspace*{5pt}} 
\setcounter{secnumdepth}{5}

\makeatletter 
\def\subsubsection{\@startsection{subsubsection}{3}{10pt}{-1.25ex plus -1ex minus -.1ex}{0ex plus 0ex}{\normalsize\bf}} 
\def\paragraph{\@startsection{paragraph}{4}{10pt}{-1.25ex plus -1ex minus -.1ex}{0ex plus 0ex}{\normalsize\textit}} 
\renewcommand\@biblabel[1]{#1}            
\renewcommand\@makefntext[1]%
{\noindent\makebox[0pt][r]{\@thefnmark\,}#1}
\makeatother 
\renewcommand{\figurename}{\small{Fig.}~}
\sectionfont{\large}
\subsectionfont{\normalsize} 

\fancyfoot{}
\fancyhead{}
\renewcommand{\headrulewidth}{1pt} 
\renewcommand{\footrulewidth}{1pt}
\setlength{\arrayrulewidth}{1pt}
\setlength{\columnsep}{6.5mm}
\setlength\bibsep{1pt}

\setlength{\textfloatsep}{10pt plus 1.0pt minus 2.0pt}

\twocolumn[
  \begin{@twocolumnfalse}
\noindent\LARGE{\textbf{Light scattering study of the ``pseudo-layer" compression elastic constant in a twist-bend nematic liquid crystal\\$^\dag$}}
\vspace{0.6cm}

\noindent\large Z. Parsouzi$^{1}$, Shokir A. Pardaev$^{1}$, C. Welch$^{2}$, Z. Ahmed$^{2}$, G. H. Mehl$^{2}$, A. R. Baldwin$^{1}$, J. T. Gleeson$^{1}$, O. D. Lavrentovich$^{3}$, D. W. Allender$^{1}$, J. V. Selinger$^{3}$, A. Jakli$^{3}$, and S. Sprunt$^{1 \ast}$
\vspace{0.5cm}

\noindent\textit{\small{\textbf{Received Xth XXXXXXXXXX 20XX, Accepted Xth XXXXXXXXX 20XX\newline
First published on the web Xth XXXXXXXXXX 200X}}}

\noindent \textbf{\small{DOI: 10.1039/b000000x}}
\vspace{0.6cm}

\noindent \normalsize{The nematic twist-bend (TB) phase, exhibited by certain achiral thermotropic liquid crystalline (LC) dimers, features a nanometer-scale, heliconical rotation of the average molecular long axis (director) with equally probable left- and right-handed domains. On meso to macroscopic scales, the TB phase may be considered as a stack of equivalent slabs or ``pseudo-layers", each one helical pitch in thickness. The long wavelength fluctuation modes should then be analogous to those of a smectic-A phase, and in particular the hydrodynamic mode combining ``layer" compression and bending ought to be characterized by an effective layer compression elastic constant $B_{eff}$ and average director splay constant $K_1^{eff}$. The magnitude of $K_1^{eff}$ is expected to be similar to the splay constant of an ordinary nematic LC, but due to the absence of a true mass density wave, $B_{eff}$ could differ substantially from the typical value of $\sim 10^6$~Pa in a conventional smectic-A. Here we report the results of a dynamic light scattering study, which confirms the ``pseudo-layer" structure of the TB phase with $B_{eff}$ in the range $10^3-10^4$~Pa. We show additionally that the temperature dependence of $B_{eff}$ at the TB to nematic transition is accurately described by a coarse-grained free energy density, which is based on a Landau-deGennes expansion in terms of a heli-polar order parameter that characterizes the TB state and is linearly coupled to bend distortion of the director.}     
\vspace{0.5cm}
 \end{@twocolumnfalse}
  ]
\footnotetext{\textit{$^{1}$~Department of Physics, Kent State University, Kent, Ohio 44242, USA\\}}
\footnotetext{\textit{$^{2}$~Department of Chemistry, University of Hull, Hull HU6 7RX, UK \\}}
\footnotetext{\textit{$^{3}$~Chemical Physics Interdisciplinary Program and Liquid Crystal Institute, Kent State University, Kent, Ohio 44242, USA\\}}

\section{Introduction}
Thermotropic liquid crystals (LCs) exemplify partial ordering in condensed matter; the panoply of distinct phases grows ever richer, challenging both experiment and theory alike to uncover and explain subtleties in the basic ordering mechanisms and properties across various length and time scales. The recently discovered twist-bend (TB) nematic phase \cite{Cestari_PRE, Borshch, Chen_PNAS, Chen_PRE, CMeyer1, Gorecka_LC} is especially remarkable in that it exhibits a molecular scale periodicity even in the absence of a periodic variation in mass density -- that is, purely in the context of orientational (nematic) order. The basis for this is believed to be the bent conformation of odd-membered LC dimers (Fig.~1) that usually form the TB phase: The bent shape promotes a structural bend, which can be accommodated without defects provided the molecular orientation also twists. The resulting heliconical structure (Fig.~1) has a notably short pitch ($t_0 \simeq 10$~nm, or a few molecular lengths) and a fairly small cone angle $\beta \simeq 10^\circ$ \cite{Borshch, CMeyer1, CMeyer2}. Typically, the dimers are achiral, and domains of left and right-handed helicity coexist.

Various theories have been put forth to explain the formation of the TB phase from a higher-temperature, uniform uniaxial nematic state. These include a theory in which the nematic bend elastic constant becomes negative below the transition temperature \cite{CMeyer2, Dozov_TB, Lelidis}, inducing a spontaneous bend that is stabilized by twist and by a positive higher-order elastic term, and theories that introduce a vector order parameter \cite{Shamid_PRE, Kats}, e.g., a polarization field, that becomes non-zero in the TB phase, and winds helically with the same nanoscale pitch as the molecular orientation to which it is coupled (Fig.~1). The latter build upon an original suggestion by Meyer \cite{BMeyer}. These models account for the heliconical microscopic structure of the TB phase via appropriate Landau-deGennes expansions of the free energy in terms of the nematic director field (locally-averaged molecular long axis), $\hat{\mathbf{n}}$, and a polarization (or similar) vector field, which we shall label $\mathbf{p}$ and take to be dimensionless -- e.g., by normalizing to a suitable low-temperature value.

Another way to view the TB phase, which would be valid on length scales long compared to the pitch, is as a phase whose optical, electrical, and mechanical properties are qualitatively similar to those of a smectic-A LC or, perhaps more appropriately given the handedness of the heliconical domains, a chiral smectic-A \cite{CMeyer3}. In this picture, slabs of the TB phase with thickness equal to one pitch are treated as smectic ``pseudo-layers" (meaning layers not delineated by a mass density wave). In addition to the Frank elastic constants of the nematic phase, two elastic moduli -- one corresponding to pseudo-layer compression ($B_{eff}$) and the other ($D_{eff}$) penalizing angular deviation of the average director from the pseudo-layer normal (or pitch axis) -- are needed to describe long wavelength distortions. The label $eff$ distinguishes the case of ``pseudo"-layering from a layering associated with the usual smectic mass density wave.

\begin{figure}[tbp]
\centering
\includegraphics[width=0.9\columnwidth]{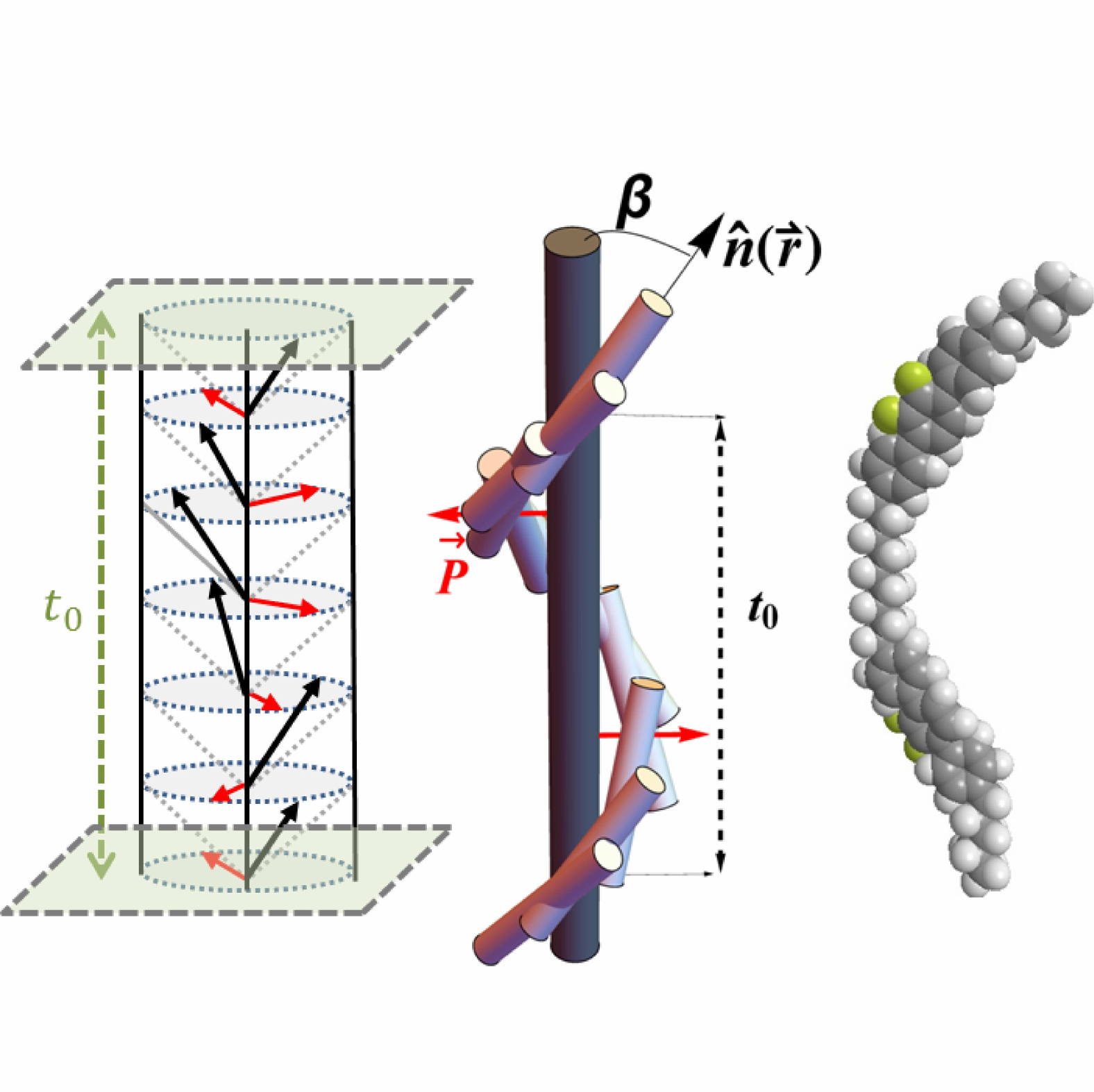}
\caption{Left and middle: Schematic views of the heliconical molecular organization in the nematic twist-bend (TB) phase. Cylinders in the middle figure represent individual molecules. The dark arrows in the left figure represent the orientation of the local molecular long axis (or heliconical director $\hat{\mathbf{n}}$), which is nonpolar. Red arrows indicate a helically modulated polar vector ($\mathbf{p}$), which represents a shape or electric polarization arising from the bent conformation of a dimer that contains an odd-numbered $CH_2$ linkage between the two aromatic core groups, such as in the dimer depicted on the right. The indicated planes, separated by one pitch length ($t_0$), define a slab-like ``pseudo-layer" of the heliconical structure.}
\end{figure}

The two theoretical approaches can be connected by a coarse-graining analysis \cite{CMeyer3,Parsouzi_PRX} of the ``microscopic" Landau-deGennes models. This analysis, which is similar to the coarse-graining of the helical structure of the cholesteric phase (where $\beta = 90^\circ$) \cite{Lubensky_PRA, Radzihovsky}, yields specific predictions for the relation between ``macroscopic" elasticities $B_{eff}$, $D_{eff}$ and the ``microscopic" parameters $q_0$ (the pitch wavenumber) and $\beta$, plus the ``bare" values of the Frank elastic constants in the nematic phase. It thereby facilitates a vital test of theory, as most experiments are conducted on length scales much larger than the nanoscale pitch.

In this paper, we report a dynamic light scattering study of the hydrodynamic fluctuation mode that combines pseudo-layer bending and compression in the TB phase. We deduce values of the compression elastic constant, $B_{eff}$, in the range $10^3-10^4$~Pa, or $\sim 10^2-10^3$ times lower than in the case of a true smectic-A mass density wave. This range agrees with estimates made from high-field magnetic birefringence measurements \cite{Challa} and rheometry \cite{Salili} on different TB materials, but contrasts with a recent report, which utilizes a different technique applied to yet other TB-forming compounds \cite{Gorecka2, Vaupotic} and obtains $B_{eff}$ in the range $\sim 10^6$~Pa of an ordinary smectic-A LC. Thus, we find $B_{eff}$ in the TB phase to be comparable to values of $\sim 10^4$~Pa reported for a tilted smectic (smectic-C) phase below the transition to the smectic-A phase \cite{SmC}, where layer compression can be accommodated by molecular tilt.

Our experimental results for the dispersion and temperature dependence of the hydrodynamic fluctuation mode validate the ``pseudo-layer" description and quantitatively support a Landau-deGennes theory of the nematic to TB transition, which invokes a polarization field as the primary order parameter. Additionally, they complement our recent study \cite{Parsouzi_PRX} of nonhydrodynamic modes (and elastic constant $D_{eff}$) in the TB phase.

\section{Theoretical Background}

In the uniform nematic phase, above the transition to the TB phase (temperature $T = T_{TB}$), light scattering probes the conventional, overdamped nematic director modes -- namely, the ``splay-bend" mode (mode 1) and the ``twist-bend" mode (mode 2) -- with scattered light intensities and relaxation rates given by \cite{deGennes},  
\begin{equation}
I_1^{N} \propto \frac{\epsilon_a^2 k_B T G_1}{K_1 q_\perp^2 + K_3 q_z^2}~~,~~\Gamma_1^{N} = \frac{K_1 q_\perp^2 + K_3 q_z^2}{\eta_1^{N}(\hat{\mathbf{q}})},
\end{equation}
\begin{equation}
I_2^{N} \propto \frac{\epsilon_a^2 k_B T G_2}{K_2 q_\perp^2 + K_3 q_z^2}~~,~~\Gamma_2^{N} = \frac{K_2 q_\perp^2 + K_3 q_z^2}{\eta_2^{N}(\hat{\mathbf{q}})}.
\end{equation}
Here $K_i$ ($i=1-3$) are the Frank elastic constants for splay, twist, and bend distortions of $\hat{\mathbf{n}}$, $T$ is the absolute temperature, $G_1$ and $G_2$ are optical factors determined by polarization and geometry-dependent selection rules, and $\mathbf{q} = (\mathbf{q}_\perp, q_z)$ is the fluctuation wavevector (with $z$ being the direction of the equilibrium director). The parameters $\eta_{1,2}^N (\hat{\mathbf{q}})$ are phenomenological viscosities, which may be expressed in terms of more fundamental nematic viscosity coefficients and the ratio $q_z/q_\perp$. As we will mainly be concerned with mode 1, we only give the expression for $\eta_1^N (\hat{\mathbf{q}})$ \cite{Viscosity_book}:
\begin{equation}
\eta_1^N (\hat{\mathbf{q}}) = \gamma_1 - \frac{(\alpha_3 - \alpha_2 q_z^2/q_\perp^2)^2}{\eta_2 + (\alpha_1+\alpha_3+\alpha_4+\alpha_5) q_z^2/q_\perp^2 +\eta_1 q_z^4/q_\perp^4}
\end{equation}
[See Ref.~[23] for definitions of the various viscosity coefficients $\gamma_1$, $\alpha_i$ ($i = 1-5$), and $\eta_i$ ($i = 1,2$).]

Turning to the TB phase, and based on the analogy to a smectic-A, we expect two fluctuation modes that directly couple to the optic axis: a ``slow", hydrodynamic layer compression-bending mode (or ``undulation" mode), with scattering intensity and relaxation rate ($\Gamma$) given by,
\begin{equation}
I_1^{TB} \propto \frac{\epsilon_a^2 k_B T G_1}{B_{eff} q_z^2/q_\perp^2 + K_1^{eff} q_\perp^2}~~,~~\Gamma_1^{TB} = \frac{B_{eff} q_z^2/q_\perp^2 + K_1^{eff} q_\perp^2}{\eta_3^{TB}},
\end{equation}
and a ``fast", non-hydrodynamic layer tilting mode, with
\begin{equation}
I_2^{TB} \propto \frac{\epsilon_a^2 k_B T G_2}{D_{eff}}~~,~~\Gamma_2^{TB} = \frac{D_{eff}}{\eta_{tilt}^{TB}}.
\end{equation}
Here $z$ corresponds to the direction of the average pseudo-layer normal, $K_1^{eff}$ is the elastic constant for splay of the average director in the TB phase, and $\eta_3^{TB}$ is a viscosity coefficient associated with pseudo-layer sliding. Fig.~2 illustrates the pseudo-layer undulation mode in the TB phase, for the case where both layer compression and bending contribute -- i.e, both $q_\perp, q_z \neq 0$. 

Eqs.~(4) and (5) assume that $B_{eff} \gg K_3^{eff} q_\perp^2$ and $D_{eff} \gg K_2^{eff} q_\perp^2\,,\,K_3^{eff} q_z^2$, where $K_2^{eff}$ and $K_3^{eff}$ are twist and bend elastic constants of the average director in the TB phase. These conditions are normally satisfied in an ordinary smectic-A LC, except very close to the transition to the nematic phase \cite{Birecki}. As evidenced in the dispersion-less nature of the nonhydrodynamic tilt mode observed in the TB phase at optical wavenumbers \cite{Parsouzi_PRX}, the second condition is validated. The first condition will be checked for self-consistency in the Results and Discussion section below. Eqs.~(4) and (5) also assume $D_{eff} \gg B_{eff} q_z^2/q_\perp^2$, so that the hydrodynamic and nonhydrodynamic modes approximately decouple; we will confirm this in the same section.

Additionally, the expressions for the ``undulation" mode ($I_1^{TB}$, $\Gamma_1^{TB}$) apply in the limit of an incompressible smectic-A (uniform mass density $\rho$), with $q_z/q_\perp \lsim \mathrm{min}(1,\lambda q_\perp)$ ($\lambda \equiv \sqrt{K_1^{eff}/B_{eff}}$) and $\rho K_1^{eff} / (\eta_3^{TB})^2 \ll 1$ \cite{Yi_PRE}. The simple form for the viscosity (single parameter $\eta_3^{TB}$) results from incompressibility and, more profoundly, from taking the hydrodynamic limit, where the ``slow" degree of freedom (hydrodynamic variable) is the pseudo-layer displacement and not rotations of the average director. From the coarse-graining models of the TB phase \cite{CMeyer3,Parsouzi_PRX}, $K_1^{eff} \approx K_1$ for small cone angle $\beta$. Then given that $K_1$ is the same order as for ordinary calamitic LCs, while $B_{eff}$ is smaller (according to our findings on the TB material studied here) and the viscosities are typically higher in the TB phase, each of the above additional conditions is met. 

\begin{figure}[tbp]
\centering
\includegraphics[width=0.9\columnwidth]{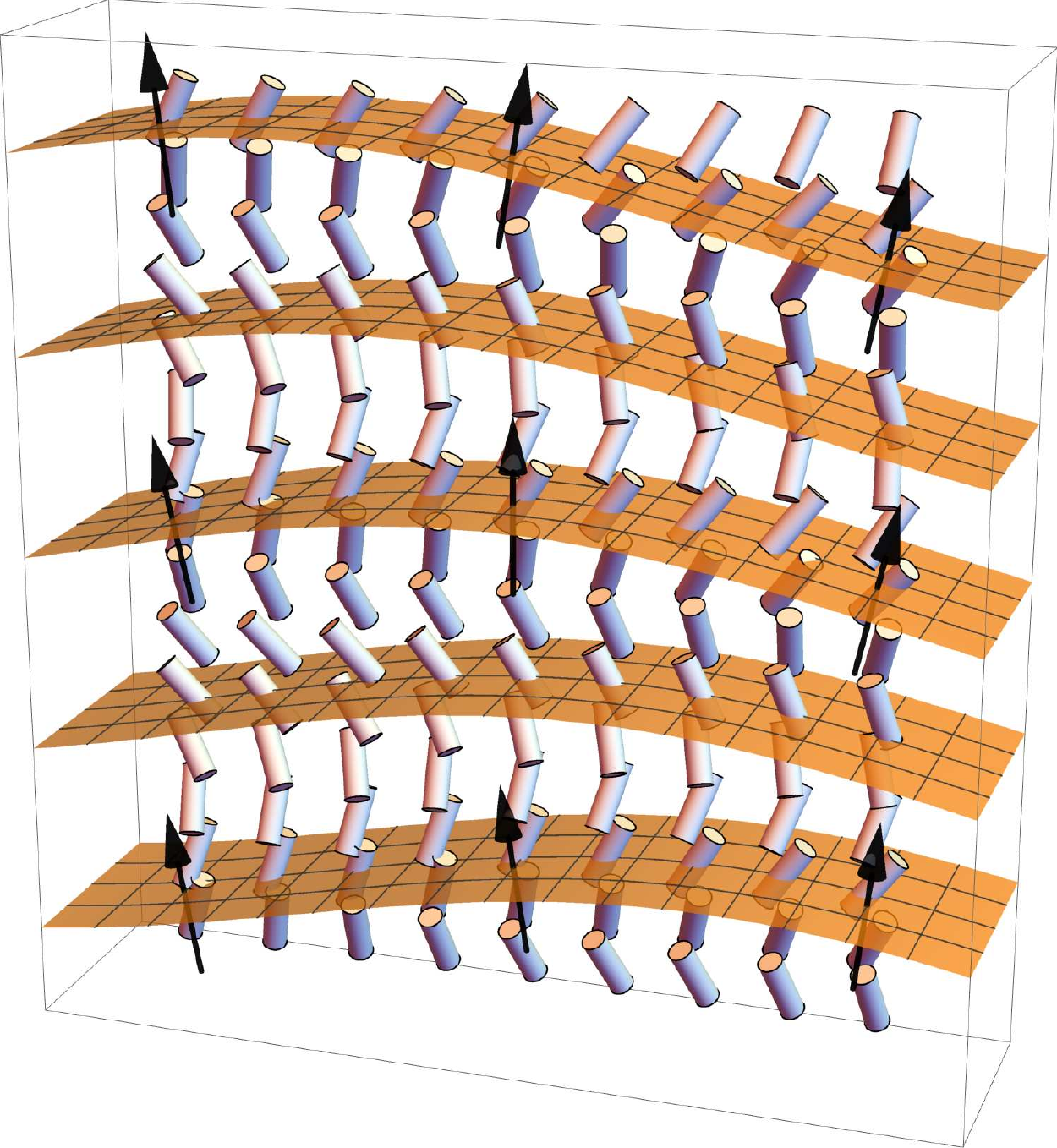}
\caption{Simulation of the pseudo-layer undulation mode in the TB phase, when $q_\perp$ and $q_z$ are both nonzero. The dark arrows represent the pseudo-layer normal (and average director).}
\end{figure}

In order to isolate $B_{eff}$ and check the wavevector dependence in Eq.~(4), we require a light scattering geometry where $I_1^{TB} \gg I_2^{TB}$, and also the capability to vary the ratio $q_z/q_\perp$. Referring to the scattering geometry in Fig.~3, with average $\hat{\mathbf{n}}$ perpendicular to the scattering plane and fixed incident angle $\theta_i = 0^\circ$, it is possible to choose a value $\theta_m$ (the so-called ``magic" angle) for the scattering angle $\theta_s$, such that $G_2 = 0$ \cite{Schaetzing}. For the present work, we used available optical birefringence data \cite{Borshch} to estimate $\theta_m = 40^\circ$ in the middle of the TB range studied.

Rocking $\hat{\mathbf{n}}$ by an angle $\chi$ off the normal to the scattering plane (see Fig.~3) then allows one to vary $q_z/q_\perp$ away from zero ($q_z = 0$ when $\chi = 0^\circ$), while introducing a minimal contribution from $I_2^{TB}$. 
In fact, since we know the magnitude of $\Gamma_2^{TB}$ for the material studied from our previous work \cite{Parsouzi_PRX}, we can verify that the contribution of $I_2^{TB}$ is negligible by the absence of a decay associated with the layer tilt mode in the measured time correlation function of the scattered light intensity.

For the fixed incident and scattered polarizations used in our experiment, the scattered light collected is a mixture of ordinary and extraordinary waves when $\chi \neq 0^\circ$. In principle, this introduces a small spread in the scattering vector (and fluctuation wavevector) $\mathbf{q}$ probed. We accounted for this in our analysis of the experimental data by allowing for a slight stretching of the single exponential decay used to fit the correlation function; however, the value of the stretching exponent always remained close to 1 (i.e., $>0.9$). 

On this basis, we may obtain an expression for $\mathbf{q} (\chi,\theta_s)$ that combines the dominant, geometrical dependence on $\chi\,,\,\theta_s$ with an approximation that takes the scattered field to lowest order to be pure ordinary (refractive index $n_o$) and the incident field to be pure extraordinary (index $n_e$) -- conditions that are exact when $\chi = 0^\circ$. We then have,
\begin{subequations}
\begin{equation}
q_\perp \approx \frac{2\pi}{\lambda_0} \left[ \left( n_e - \sqrt{n_o^2-\sin^2 \theta_s} \right)^2 + \sin^2 \theta_s \cos^2\chi \right]^{1/2}  
\end{equation}
\begin{equation}
q_z \approx \frac{2\pi}{\lambda_0} \sin \theta_s \sin \chi,
\end{equation}
\end{subequations}
where the angle $\theta_s$ is measured in the lab, and $\lambda_0$ is the wavelength of light in air.

\section{Experimental Details}

The LC material studied is a 70/30 wt\% mixture of the dimer and monomer compounds shown in Fig.~3, which we abbreviate DTCm \cite{Panov}. Its phase sequence (on cooling) is: isotropic$\rightarrow$N$\rightarrow$TB$\rightarrow$crystal, with the transition to the TB phase occurring at approximately $88.25^\circ$C. The mixture DTCm was chosen for the following reasons. First, the dielectric anisotropy $\epsilon_a$, which generally decreases at the nematic to TB transition, does not decrease by much in DTCm \cite{Borshch}, and thus its temperature dependence becomes a weak, secondary factor in the behavior of the measured light scattering intensity $I$. This simplifies the connection between the temperature dependence of $B_{eff}$ and that of $I$ for the hydrodynamic fluctuation mode. Second, DTCm has been thoroughly characterized by various techniques, ranging from freeze-fracture TEM \cite{Borshch}, which directly reveals the nanoscale orientational modulation, to light scattering measurements of nonhydrodynamic modes in the TB phase \cite{Parsouzi_PRX}. 

\begin{figure}[tbp]
\centering
\includegraphics[width=0.9\columnwidth]{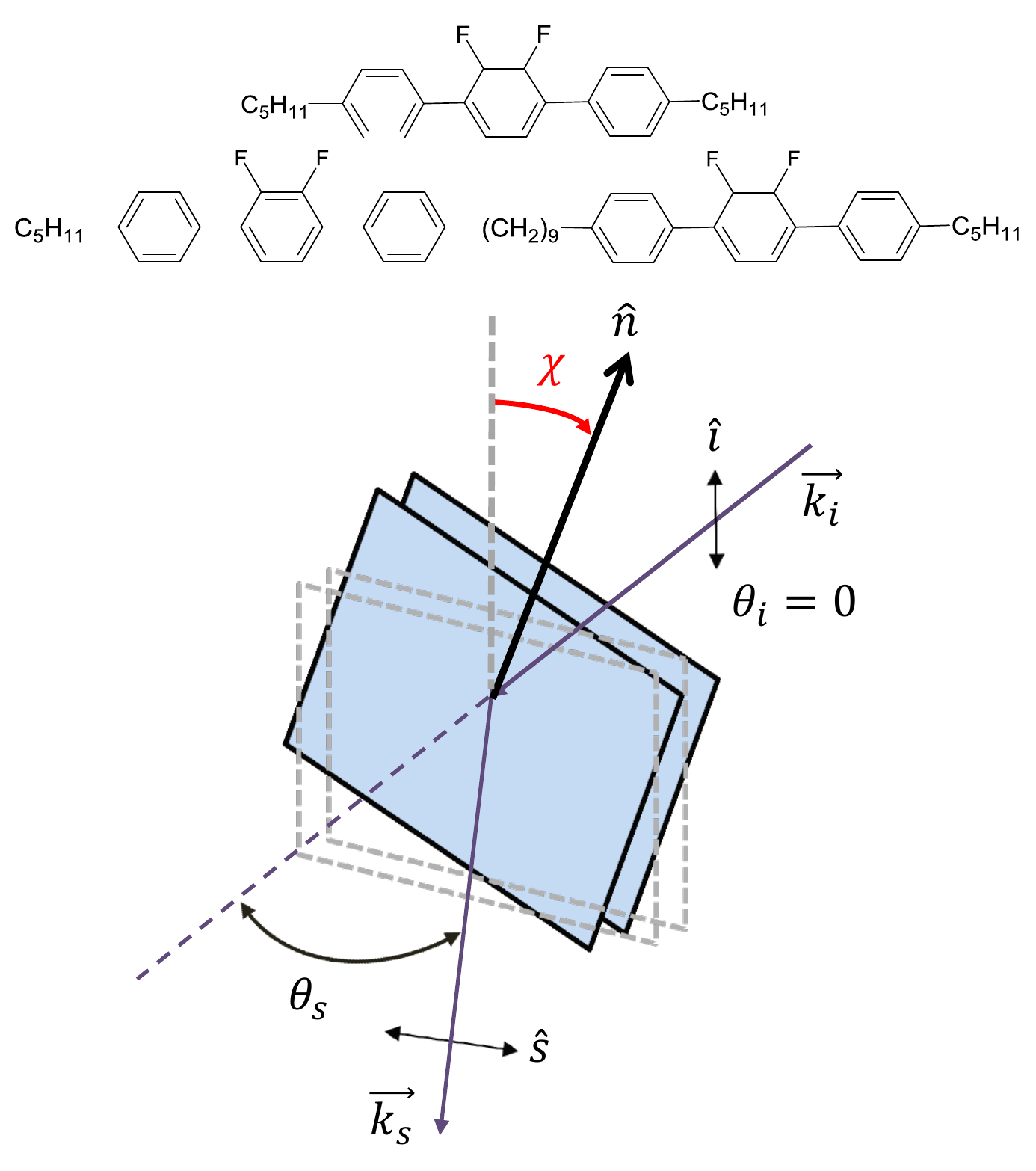}
\caption{Top: Chemical structure of the dimer and monomer components of the mixture studied. (A 3D rendering of the minimum energy conformation of the dimer is shown in Fig.~1.) Bottom: Light scattering geometry, with the ``rocking" angle $\chi$ indicated. The normally incident laser light (wavevector $\mathbf{k}_i$, incident angle $\theta_i$) is polarized vertical to the scattering plane (and parallel to the average director $\hat{\mathbf{n}}$ when $\chi = 0^\circ$). Horizontally-polarized scattered light (wavevector $\mathbf{k}_s$ is collected at angle $\theta_s$. The fluctuation wavevector probed is $\mathbf{q} = \mathbf{k}_s - \mathbf{k}_i$.} 
\end{figure}

Homogeneous planar-aligned nematic samples of DTCm were prepared using commercial cells (EHC, Japan) with $4\mu$m nominal spacing between flat optical substrates that have rubbed polyimide alignment layers. The sample cells were placed in a microscope hot stage, temperature-regulated to $0.002^\circ$C precision and slightly modified for light scattering studies. The hot stage was mounted on a three circle goniometer. Two coplanar, horizontal circles provided adjustment of incident and scattering angles ($\theta_i$ and $\theta_s$), and the third circle, mounted vertically, enabled the nematic director (or equilibrium pitch axis in the TB phase) to be continuously rotated (through angle $\chi$) between parallel and perpendicular orientations with respect to the scattering plane (Fig.~3). Separate $xy$ micro-positioning stages allowed the rotation axis of the third circle to be positioned precisely in coincidence with the normally-incident, vertically-polarized laser beam ($532$~nm wavelength, $\sim 5$~mW incident power), and to vary the position of the illuminated volume in the sample. A long distance polarizing microscope was situated in the scattering plane and used to monitor both the sample texture and the precise position of the beam on the sample during the light scattering measurements.

Horizontally polarized scattered light was collected at various $\theta_s$ and $\chi$, and the intensity-intensity time correlation function was computed and recorded on a homemade digital electronic correlator. Fig.~4 displays representative, normalized light scattering correlation functions, taken at two values of angle $\chi$ ($0^\circ$ and $30^\circ$) for fixed scattering angle $\theta_s = \theta_m = 40^\circ$, in the TB phase ($T - T_{TB} = -2.6^\circ$C) of DTCm. The solid lines through the data represent fits to a slightly stretched, single exponential decay. Examples of the texture and position of the illuminated volume in the sample for two values of $\chi$ are also displayed. A weak stripe texture is evident in the image for $\chi = 0^\circ$; this is most likely due to a slight pseudo-layer distortion or ``buckling" near the cell surfaces. We carefully positioned the illuminated volume to minimize static light scattering from the stripes and to maintain the signal/background ratio of the correlation function above 90\%. For measurements in the TB phase, the sample was very slowly cooled through $T_{TB}$. Again, the choice of material proved advantageous, as the cone angle for DTCm varies slowly below $T_{TB}$, minimizing pseudo-layer strains at the surface.

\begin{figure}[tpb]
\centering
\includegraphics[width=0.9\columnwidth]{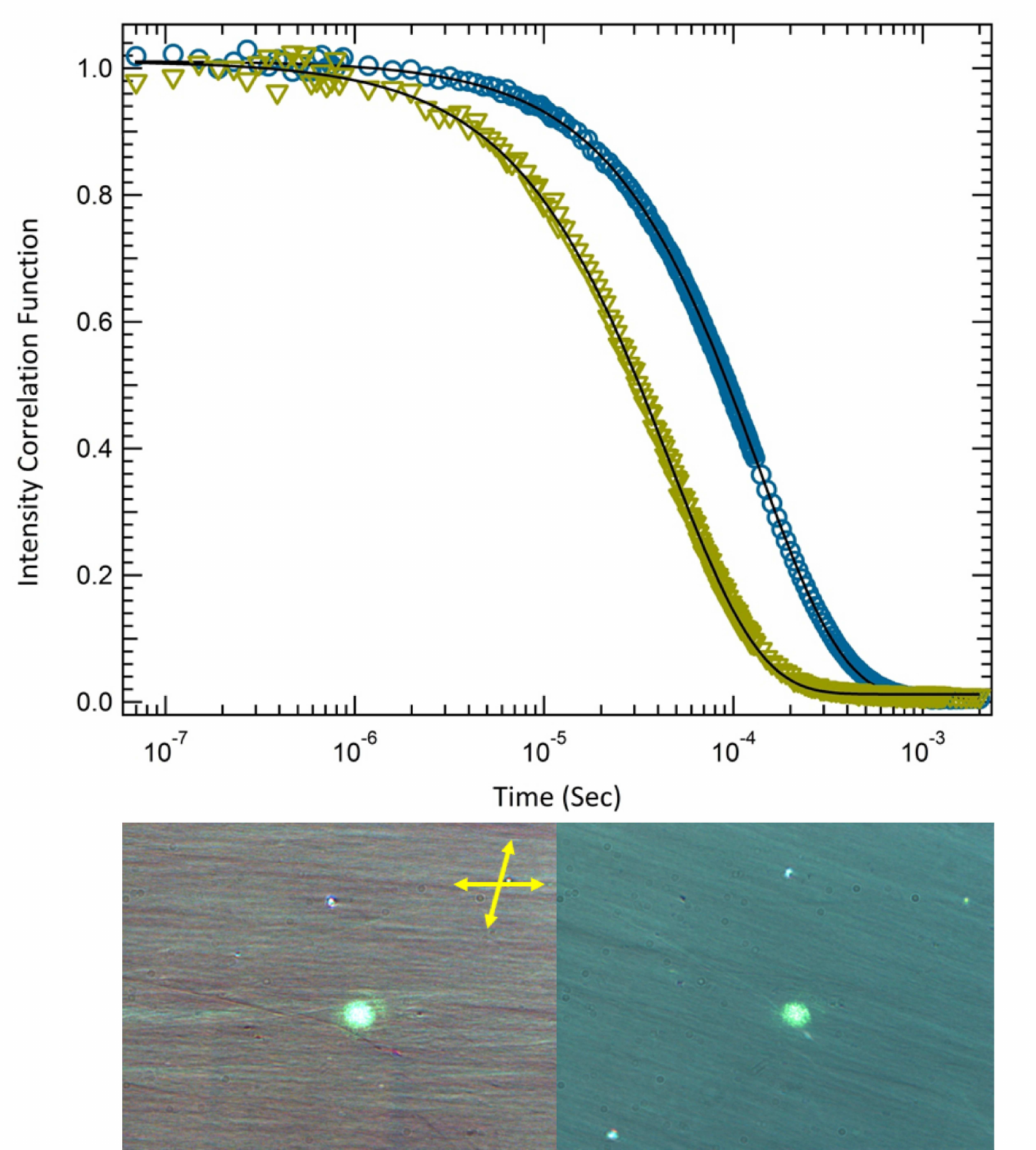}
\caption{Top: Typical light scattering correlation functions obtained in the TB phase at $T-T_{TB} = -2.6^\circ$C and for normal incidence, scattering angle $\theta_s = 40^\circ$, and rocking angles $\chi = 0^\circ$ (right plot) or $30^\circ$ (left plot). Solid lines are fits to a slightly stretched single exponential decay. Bottom: Textures of the TB phase recorded by polarizing microscopy at $T_{TB}-T = -0.6^\circ$C and for angle $\chi = 0^\circ$ (left) and $16^\circ$ (right). The position of the scattering volume is also recorded, allowing us to confirm no translation of the illuminated volume when the sample is rocked. The weak stripe texture visible for $\chi = 0^\circ$ is probably due to pseudo-layer shrinkage at the cell surfaces; it caused no significant static scattering.}
\end{figure}

\section{Results and Discussion}

The main results obtained from analysis of the correlation data, or from measurements of the scattered intensity (normalized to incident laser power), are contained in Figs.~5-7.

\begin{figure}[tbp]
\centering
\includegraphics[width=1\columnwidth]{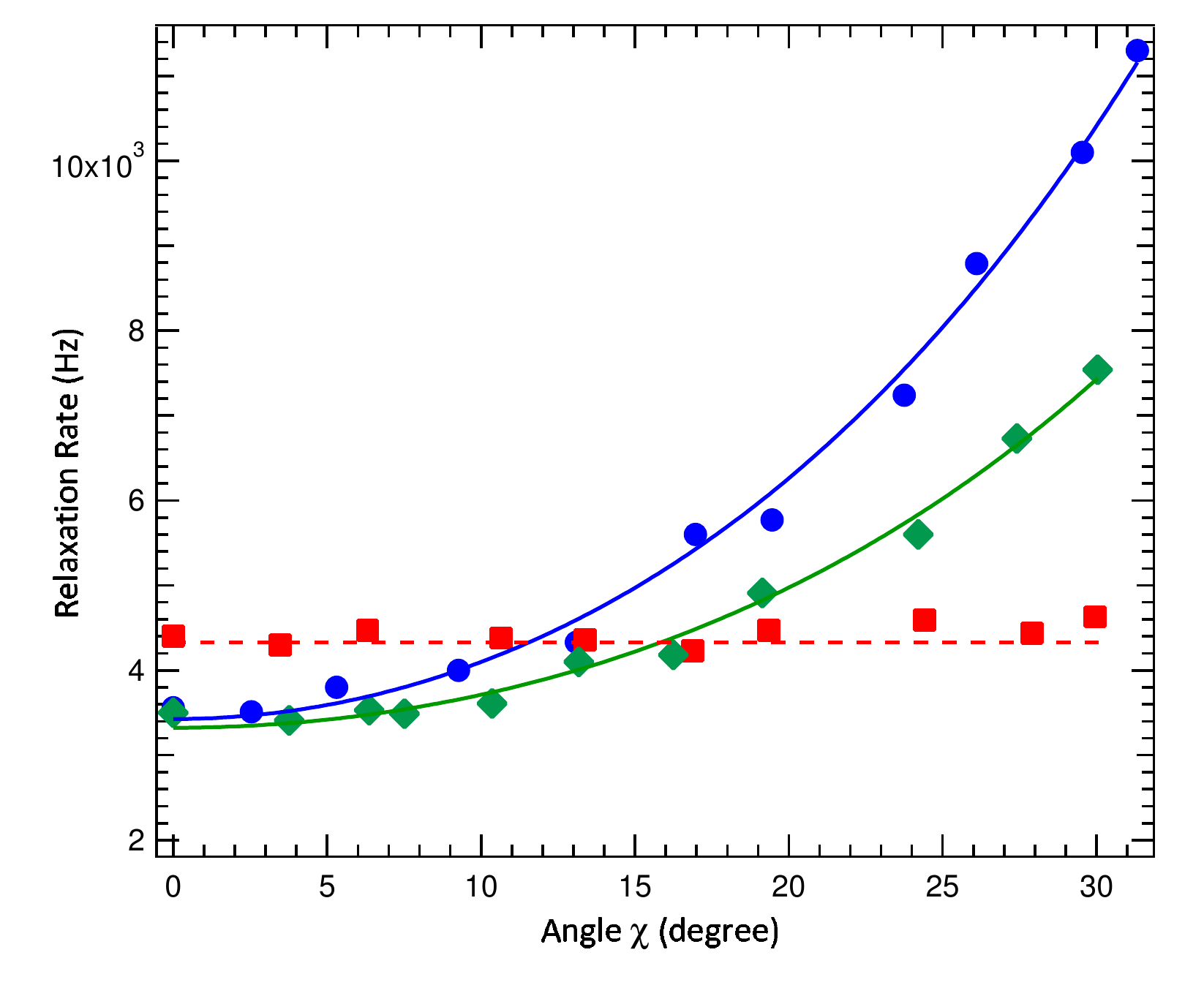}
\caption{Dependence of the relaxation rate $\Gamma_1^{TB}$ of the pseudo-layer undulation mode on the rocking angle $\chi$ for two temperatures, $T-T_{TB} = -1.4$ (green diamonds) and $-2.6^\circ$C (blue circles), in the TB phase of DTCm, and for scattering angle $\theta_s = \theta_m = 40^\circ$. The solid lines are fits to the combination of Eqs.~(4) and (6), as described in the text. The red squares represent data for $\Gamma_1^N$ at $T-T_{TB} = 1.9^\circ$C in the nematic phase, and the dashed line represents a constant value.} 
\end{figure}

Fig.~5 presents the relaxation rate $\Gamma_1^{TB}$ of the pseudo-layer undulation mode as a function of rocking angle $\chi$ for $\theta_s = \theta_m = 40^\circ$ at two temperatures $T-T_{TB} = -1.4^\circ$ and $-2.6^\circ$C in the TB phase, and $\Gamma_1^N$ vs $\chi$ for the same $\theta_s$ at $T-T_{TB} = 1.9^\circ$C in the nematic state. Below $T_{TB}$, $\Gamma_1^{TB}$ clearly has the behavior expected from Eq.~(4); it increases with $q_z$, which depends on $\chi$ according to Eq.~(6b). The solid curves are fits of the data to the combination of Eq.~(4) for $\Gamma_1^{TB}$ and Eqs.~(6) for $q_\perp$ and $q_z$. The index anisotropy $n_e - n_o$ is known for DTCm as a function of temperature \cite{Borshch}. If we take $n_o \simeq 1.5$ (higher precision does not significantly affect the results of our analysis), the quantity $\left( n_e - \sqrt{n_o^2-\sin^2 \theta_s} \right)^2$ in the expression for $q_\perp$ can be estimated as $0.09$ when $\theta_s = 40^\circ$. This leaves two adjustable parameters in our fit for $\Gamma_1^{TB}$, namely $B_{eff}/\eta_3^{TB}$ and $K_1^{eff}/\eta_3^{TB}$, whose ratio gives $B_{eff}/K_1^{eff}$.

The fit yields $B_{eff}/K_1^{eff} = 3.9 \times 10^{14}$~m$^{-2}$ and $6.1 \times 10^{14}$~m$^{-2}$ for $T - T_{TB} = -1.4$ and $-2.6^\circ$C, respectively. Then we can obtain $B_{eff}$ from an estimate of $K_1^{eff}$. In both a conventional smectic-A and in the "pseudo"-layer model of a TB phase with small cone angle $\beta$, $K_1^{eff}$ is comparable to the nematic splay constant $K_1$. In the mixture we study, the measured $K_1$ varies from $\sim 2$~pN (close to the nematic to isotropic transition) to $33$~pN (near $T_{TB}$) \cite{Borshch}. Taking $K_1 = 15$~pN, we find $B_{eff} = 5.9 \times 10^3$~Pa and $9.2 \times 10^3$~Pa at the two temperatures $T-T_{TB} = -1.4^\circ$C and $-2.6^\circ$C. These values would increase by a factor of $\sim 2$, if we used the value of $K_1$ just above $T_{TB}$.

We can compare our experimentally deduced values for $B_{eff}$ with the predictions of the coarse-grained theories of the TB phase in Refs.~13,14, which both predict the order of magnitude $B_{eff} \approx K_3 q_0^2 \beta^2$. Then taking previously measured values $K_3 = 2 \times 10^{-12}$~N (characteristic of the nematic phase of DTCm), $\beta = 5.5^\circ$ (for $T-T_{TB} \simeq -2^\circ$C) and $q_0 = 2 \pi/t_0$ with pitch $t_0 = 9.3$~nm for DTCm (in the TB phase) \cite{Borshch}, we get $B_{eff} \approx 8.4 \times 10^3$~Pa, which falls in the same range as our experimental values.

We may now check the assumption $B_{eff} \gg K_3^{eff} q_\perp^2$ made in Eq.~(4) (Theoretical background section). For small $\beta$, the coarse-graining models give $K_3^{eff} = K_3 + O (\beta^2)$ \cite{CMeyer3,ShSelUnpub}. Then using $K_3 = 2 \time 10^{-12}$~N and max $q_\perp = 7.6 \times 10^6$~m$^{-1}$, we get $K_3^{eff} q_\perp^2 \approx 120$~Pa, which is much smaller than $B_{eff}$ extracted from our measurements.

While our experimental values for $B_{eff}$ are in agreement with the coarse-graining theories of the TB state, they differ markedly from recently reported experimental results \cite{Gorecka2, Vaupotic} based on a different technique and on different TB materials. In particular, values of $B_{eff}$ were reported in the range $10^6-10^7$~Pa for the pure dimer CB7CB. These values are typical of a true smectic-layered phase (i.e., a phase with a 1D mass density wave, as opposed to purely orientational modulation), and are much larger than the values we obtain for DTCm. 

In fact, we would need $K_1^{eff} \sim 10^3 K_1$ -- which would imply a scattering intensity several orders of magnitude lower than we observe for $\chi = 0^\circ$ in the TB phase -- to produce $B_{eff} \sim 10^6$~Pa. The discrepancy in $B_{eff}$ reported for different TB materials clearly highlights the need for additional studies with different techniques applied to common samples. However, it is probably essential that all measurements be performed on well-aligned TB samples with minimal (or no) stripe texture -- that condition is certainly important for light scattering. Since the stripes in planar-aligned cells of CB7CB are severe (presumably due to the rapid increase in cone angle with decreasing temperature below $T_{TB}$ \cite{CMeyer2}), and not easily mitigated by laboratory-scale applied fields, we have so far not been able to perform meaningful measurements on this material in the TB phase.

Fig.~5 also shows the $\chi$ dependence of $\Gamma_1$ in the nematic phase. According to the numerator of the expression for $\Gamma_1^N$ in Eq.~(1), and recalling that $K_1 \gg K_3$ in DTCm \cite{Borshch}, the relaxation rate should decrease slightly with increasing $\chi$; this is due to the $\cos^2 \chi$ factor in the expression for $q_\perp$ in Eq.~(6a). Our data show, however, that $\Gamma_1^N$ remains relatively flat. The reason for this could be an offsetting effect due to the $\hat{\mathbf{q}}$ dependent orientational viscosity $\eta_1^N (\hat{\mathbf{q}})$, Eq.~(3). As $\chi$ increases from zero at the ``magic" scattering angle $\theta_s = \theta_m$, and according to Eq.~(3), $\eta_1^N (\hat{\mathbf{q}})$ begins to cross over from $\eta_{splay} = \gamma_1 - \alpha_3^2/\eta_2$ (when $q_z = 0$) to $\eta_{bend} = \gamma_1 - \alpha_2^2/\eta_1$ (when $q_\perp = 0$). Since in typical nematics $\eta_{splay} \simeq (4-5) \, \eta_{bend}$ \cite{G.Chen}, a decrease in the denominator of the expression for $\Gamma_1^N$ with $\chi$ could cancel the decrease in the numerator, resulting in an essentially constant value as we observe from the experimental data.

\begin{figure}[tbp]
\centering
\includegraphics[width=1\columnwidth]{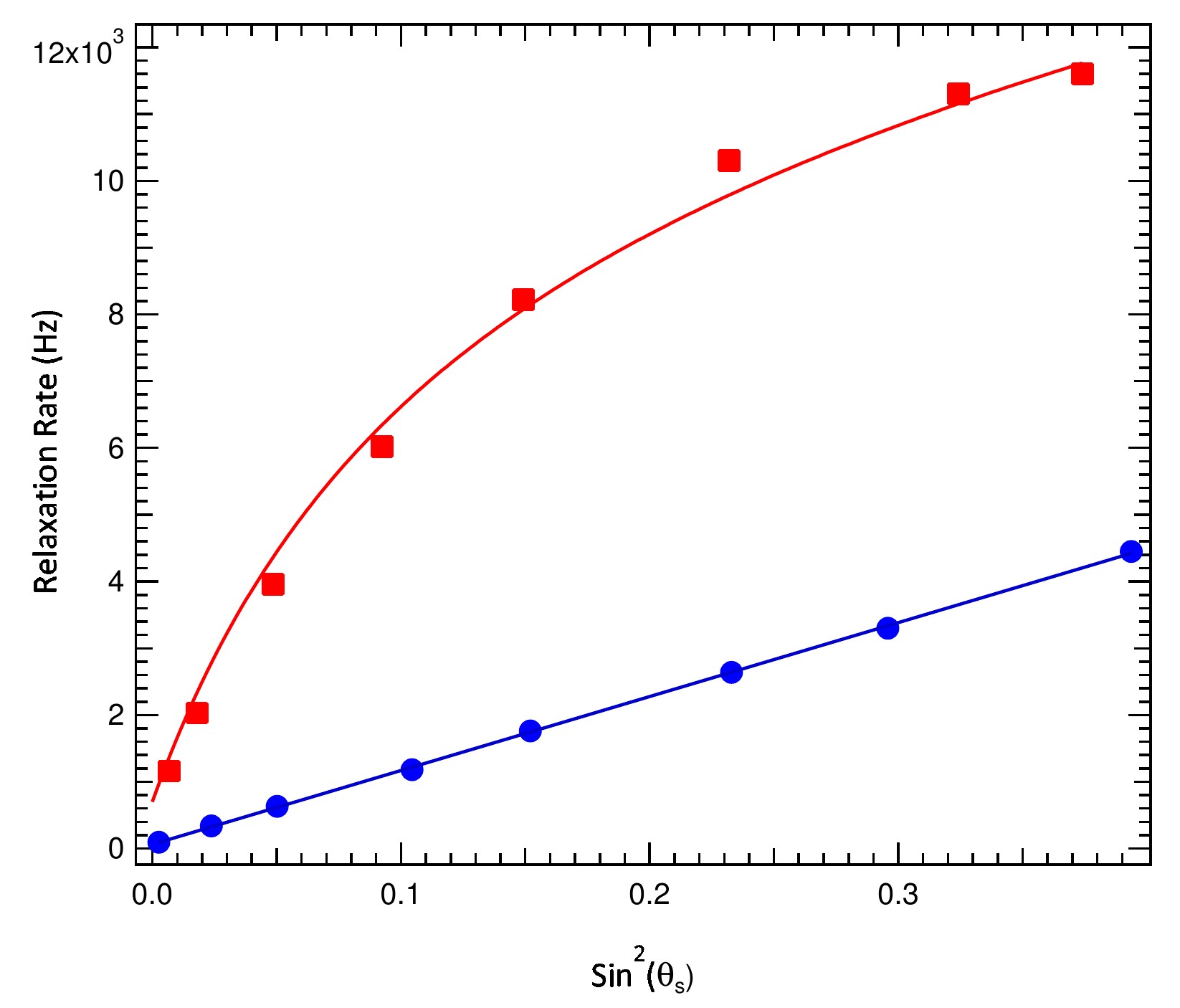}
\caption{Relaxation rate $\Gamma_1$ versus $\sin^2\theta_s$ for rocking angle $\chi = 30^\circ$ in the TB ($T-T_{TB} = -2.5^\circ$C, red squares) and nematic ($T-T_{TB} = 4.5^\circ$C, blue circles) phases of DTCm. The solid lines are fits to the combination of Eqs. (1) or (4) and (6), as described in the text.} 
\end{figure}

Let us next turn to the dependence of relaxation rate $\Gamma_1$ on the magnitude of the scattering vector for fixed $\chi = 30^\circ$. Typical data in the nematic and TB phases are displayed in Fig.~6. These data allow us to test, in particular, the dispersion relation for the pseudo-layer undulation mode in Eq.~(4). According to Eqs.~6, the quantity $\sin^2 \theta_s$ (horizontal axis in Fig.~6) controls the magnitudes of $q_\perp$ and $q_z$. In the nematic phase ($T>T_{TB}$), $\Gamma_1^N (\theta_s) \sim ( n_e - n_o )^2 + \left[ \cos^2\chi + (K_3/K_1) \sin^2 \chi \right] \sin^2 \theta_s$ from Eqs.~(1) and (6) and using $n_o^2 \gg \sin^2\theta_s$. Since $(n_e - n_o)^2 \simeq 0.025 \ll \cos^2 \chi = 0.75$, we expect $\Gamma_1^N$ to be linear in $\sin^2 \theta_s$ with a very small intercept, and the fit in Fig.~6 confirms this.

On the other hand, in the TB phase, the dependence of $\Gamma_1^{TB}$ on $\sin^2\theta_s$ is expected to be nonlinear due to the $q_z^2/q_\perp^2$ term in Eq.~(4). For small $\theta_s$, Eq.~(6) gives $q_z^2/q_\perp^2 \sim \sin^2 \theta_s$, while at large $\theta_s$, the ratio saturates at a value of $\tan^2 \chi$. The behavior of the data for $\Gamma_1^{TB}$ in Fig,~6 are qualitatively consistent with this prediction. Quantitatively, we can fit the data to the expression for $\Gamma_1^{TB}$ obtained from the combination of Eqs.~(4) and (6), with the ratio $B_{eff} / K_1^{eff}$ fixed according to the calculation above (from the rocking angle scan) and with only a single variable parameter (an overall scale factor), provided we assume that the pseudo-layers are rigidly anchored at the substrate surfaces so that the minimum $q_\perp$ for the undulation mode is cut off by the finite sample thickness, $q_{\perp,\mathrm{min}} \simeq \pi / d$ ($d =$~sample thickness). Thus we replace the first term in brackets in Eq.~(6a) with the long wavelength cut-off $q_{\perp,\mathrm{min}}^2$. The result of the fit, shown as the solid red line in Fig.~6, not only is consistent with the value of $B_{eff} / K_1^{eff}$ determined from the $\chi$ scan (at essentially the same temperature, Fig.~4), but also directly supports the ``pseudo-layer" model of the TB phase, which leads directly to the $\mathbf{q}$ dependence for $\Gamma_1^{TB}$ in Eq.~(4).
 
\begin{figure}[tbp]
\centering
\includegraphics[width=1\columnwidth]{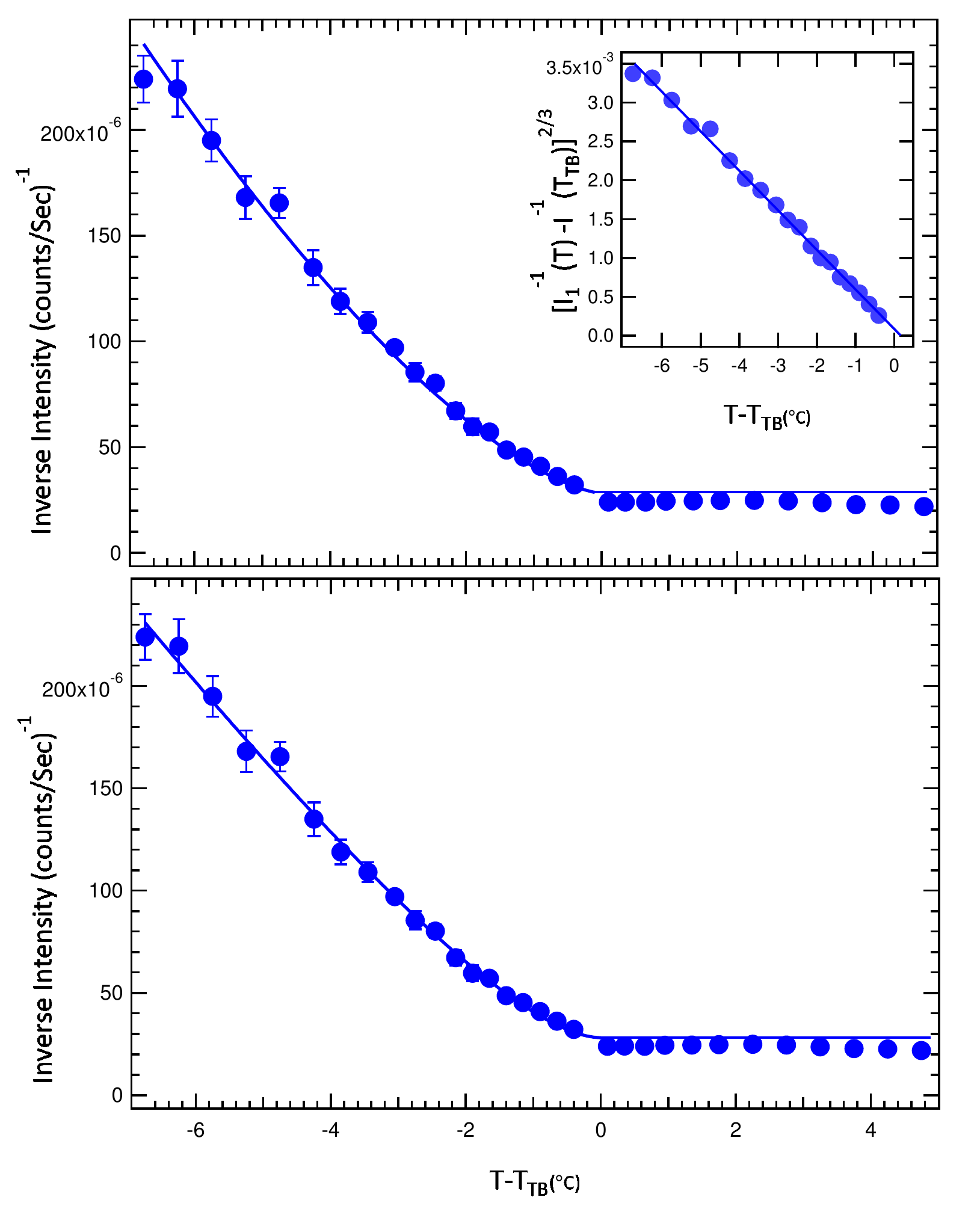}
\caption{Main panels: Temperature dependence of the inverse scattered intensity $I_1^{-1}$ from DTCm for scattering angle $\theta_s = \theta_m = 40^\circ$ and rocking angle $\chi = 30^\circ$. In the bottom panel, the solid line is a fit of the data to the theoretical expressions, Eq.~(7) and (8) of the text, describing the temperature dependence of $B_{eff}$ (which scales linearly with the temperature dependence of $I_1^{-1}$). The solid line in the top panel is a fit to the alternative $T$ dependence predicted by Eqs.~(9) and (8). The inset to the top panel shows the quantity $[I^{-1}(T) - I^{-1}(T_{TB})]^{2/3}$, calculated from the data in the TB phase, as a function of $T$ and a fit to a straight line (see discussion in text).} 
\end{figure}

In the above analysis, one might be concerned about mode 2 contaminating the scattering from mode 1 when $\theta_s$ deviates significantly from $\theta_m$ or for $\chi$ significantly off of $0^\circ$. However, in the TB phase, comparing the scattering intensities of the two modes measured in the present and our previous work \cite{Parsouzi_PRX}, we estimate that $I_1^{TB} \gsim 10 I_2^{TB}$ for all $\chi$, $\theta_s$ studied in Figs.~5 and 6. In the nematic phase, the mode 1 and 2 relaxation rates have similar dependence on $\sin^2 \theta_s$, so the result in Fig.~6 that $\Gamma^N \sim \sin^2 \theta_s$ should not change even if mode 2 contributes slightly.      

Finally, we consider the temperature dependence of $B_{eff}$. In a coarse-graining analysis, the theoretical model of the nematic to TB transition that emphasizes the role of a helical polar order parameter \cite{Shamid_PRE} makes the following predictions: Assuming that the cone angle ($\beta$) and polar order magnitude ($| \mathbf{p} |$) do not relax under pseudo-layer compression or dilation (i.e., under variations in heliconical pitch), we find \cite{Parsouzi_PRX}:
\begin{equation}
B_{eff} = \Lambda p_0 q_0 \sin \beta \cos \beta \approx \frac{\Lambda^2}{K_3} p_0^2.
\label{parsouzi_Beff}
\end{equation}
Here, $\Lambda$ is the coupling between bend distortion of the director and $\mathbf{p}$, $K_3$ is the ``bare" nematic bend elastic constant, $\beta$ is assumed to be small, and the temperature dependence of $|\mathbf{p}| = p_0$ is given by  
\begin{equation}
p_0 (T) = - \frac{3 \Lambda^2 (\kappa K_2)^{1/2}}{2 K_3^2 \nu} + \sqrt{ \frac{9 \Lambda^4 \kappa K_2}{4 K_3^4 \nu^2} + \frac{\mu_0}{\nu} (T_{TB} - T)},
\label{parsouzi_p0}
\end{equation}
where $\mu_0$ and $\nu$ are Landau coefficients.

On the other hand, if $\beta$ and $| \mathbf{p} |$ are allowed to vary by small amounts in response to small variations in pseudo-layer spacing, the scaling of $B_{eff}$ with $p_0$ to leading order changes to
\begin{equation}
B_{eff} \approx \frac{3 \sqrt{K_2 \kappa} \Lambda^2}{K_3^2} p_0^3
\end{equation}
This result is calculated by using Eqs.~(1)-(3) of Ref.~14 to obtain the free energy density of the TB phase ($F_{TB}$) as a function of $q_0$, $| \mathbf{p} |$, and $\beta$, and then treating all three variables as variational parameters: $q_0 \rightarrow q_0 + \delta q$, $| \mathbf{p} | \rightarrow p_0 + \delta p$, and $\beta \rightarrow \beta + \delta \beta$.  The variational change in $F_{TB}$, $\delta F_{TB}$, is then approximated by the Taylor series expansion out to quadratic order in $\delta q$, $\delta p$, and $\delta \beta$.  Minimizing $\delta F_{TB}$ with respect to $\delta p$ and $\delta \beta$ then gives $\delta p$ and $\delta \beta$ as proportional to $\delta q$. Substituting these values into $\delta F_{TB}$ yields $\delta F_{TB} = g (\delta q)^2 / 2$, where the factor $g$ depends on all six second derivatives of $F_{TB}$ with respect to $q_0$, $p_0$, and $\beta$. After coarse graining, one arrives at $B_{eff} = q_0^2 g$. To lowest order in $p_0$, $q_0^2 = (\Lambda^2 / K_3^2)^2 \sqrt{K_2 / \kappa} p_0$ \cite{Parsouzi_PRX}, and one also finds $g = 3 \kappa p_0^2$. Eq.~(9) then follows from these results.

Two different regimes of the temperature dependence of $p_0$ in Eq.~(8) can be distinguished by defining a cross-over temperature $T_x = T_{TB} - (9 \Lambda^4 \kappa K_2) / (4 K_3^4 \mu_0 \nu)$. For $T \ll T_x < T_{TB}$, Eq.~(8) simplifies to $p_0 \sim \sqrt{T_{TB}-T}$, and then Eqs.~(7) and (9) give, respectively, $B_{eff} \sim T_{TB}-T$ and $B_{eff} \sim (T_{TB} - T)^{3/2}$. However, for $T_x \lsim T < T_{TB}$, the temperature dependence of $p_0$ crosses over to $p_0 \sim T_{TB}-T$, and we have $B_{eff} \sim (T_{TB} - T)^2$ (from Eq.~(7)) and $B_{eff} \sim (T_{TB} - T)^3$ (from Eq.~(9)). The cubic scaling in the last expression agrees with a coarse-graining theory of the TB phase based on negative bend elasticity \cite{CMeyer3,Dozov_private}, assuming that the cone angle $\beta$ is allowed to relax under pseudo-layer compression/dilation. For $T$ sufficiently close to $T_{TB}$, this theory gives,
\begin{equation}
B_{eff} = -\frac{4}{3} K_3 q_0^2 \sin^2 \beta = -\frac{4}{27 K_2 C} K_3^3\approx \frac{4(K_3^0)^3}{27 C K_2} (T_{TB}-T)^3,\nonumber
\label{dozov_Beff}
\end{equation}
where $K_3 = K_3^0 (T - T_{TB})$ is an effective bend constant that becomes negative at $T_{TB}$, and $C>0$ is a higher order elastic constant that stabilizes the elastic free energy.

From Eq.~(4), the inverse scattered intensity from the undulation mode in the TB phase is $(I_1^{TB})^{-1} (T) \propto B_{eff} (T) q_z^2/q_\perp^2 + K_1^{eff} q_\perp^2$, where $K_1^{eff} \approx K_1$ (the ``bare" splay constant), and we neglect the weak temperature dependence of $\epsilon_a$ in the TB phase (which is valid for DTCm \cite{Borshch}). Thus, data for inverse intensity versus temperature can be fitted to the theoretical expressions above for the temperature dependence of $B_{eff}$. The solid line in the bottom panel of Fig.~7 is a fit to Eqs.~(\ref{parsouzi_Beff}--\ref{parsouzi_p0}), with three variable parameters (which are proportional to $3 \Lambda^2 (\kappa K_2 )^{1/2} / (2 K_3^2 \nu)$, $\mu_0/\nu$, and $K_1$). The fit parameters give an estimate of $T_{TB}-T_x \approx 0.7^\circ$C, which is similar to the estimate of $0.3^\circ$C obtained in our previous study of the non-hydrodynamic modes in DTCm \cite{Parsouzi_PRX}.

The top panel of Fig.~7 shows the data for $I_1^{-1}$ vs $T$ analyzed according to the alternative prediction in Eq.~(9). In this case, for $T$ sufficiently below $T_{TB}$, we expect $B_{eff}$ to scale as $(T_{TB}-T)^{3/2}$, so that $[ I_1^{-1} (T) - I_1^{-1} (T_{TB}) ]^{2/3} \sim B_{eff}^{2/3} \sim T_{TB}-T$. As shown in the inset, which plots $[ I_1^{-1} (T) - I_1^{-1} (T_{TB}) ]^{2/3}$ vs $T$, the data for $I_1^{-1} (T)$ are also consistent with the $T$ dependence predicted by Eqs.~(9) and (8). The main plot in the top panel shows the result of fitting the TB phase data to this prediction. The best fit occurs for $T_{TB} - T_x \approx 0$ -- i.e., for a much narrower cross-over region below the transition than indicated by the fit using Eqs.~(7) and (8). From our definition of $T_x$, such a narrow cross-over range would suggest that the product $\mu_0 \nu \gg \Lambda^4 \kappa K_2 / K_3^4$, though the ratio $\mu_0 / \nu$ could still have a wide range of values.

The quality of the fits in Fig.~7 to the two different predictions for the scaling of $B_{eff}$ with $T$ is fairly good, and quite comparable, over the TB range studied. Determining which scaling relation is the correct one for DTCm clearly requires a more accurate determination of the cross-over temperature $T_x$ and thus acquisition of considerably higher resolution data near $T_{TB}$. 

On the nematic side, close to $T_{TB}$, the model predicts essentially constant $I_1^{-1}$ (again ignoring small variations of $\epsilon_a$ with $T$). In both panels of Fig.~7, the model appears somewhat higher than the nematic data. This suggests an additional contribution to the experimentally measured intensity, which can be accounted for as a small contribution from mode 2 (i.e., from the hydrodynamic twist-bend mode, $I_2^N$), which is only expected to vanish when both $\theta_s = \theta_m$ and $\chi = 0^\circ$. 

\section{Conclusion}

In this work, we have presented light scattering measurements of the ``pseudo-layer" undulation mode in the twist-bend nematic phase of a material that shows minimal surface -induced distortion of the pseudo-layers (``stripe" texture) and thus allows high quality measurements. We obtained estimates of the pseudo-layer compression modulus $B_{eff}$ in the range $\sim 10^3-10^4$~Pa, confirmed the smectic-A-like dispersion relation for hydrodynamic pseudo-layer compression/bending fluctuations, and demonstrated agreement between the measured temperature dependence of $B_{eff}$ and predictions of the coarse-graining of a Landau-deGennes theory of the nematic to TB phase transition, which features a vector polarization field as the primary order parameter and invokes a linear coupling between this field and bend distortion of the director. Further experiments, conducted very close to $T_{TB}$, are necessary (at least in the material studied) to determine whether or not the pseudo-layers fluctuate ``adiabatically" with respect to the microscopic degrees of freedom (helipolar order parameter and cone angle) of the heliconical TB structure. Additionally, it would be interesting to perform similar light scattering studies on other TB-forming materials, provided the stripe texture can be effectively suppressed.     

\section*{Acknowledgements} 
This work was supported by the NSF under grants DMR-1307674 (ZP, SP, JG, AJ, and SS), DMR-1409658 (JVS), DMR-1410378 (ODL), and the EPSRC under grant EP/J004480/1 (GM and CW).\\

\end{document}